\documentclass{emulateapj}

\shorttitle{Relative orbit orientation}
\shortauthors{Tokovinin \& Latham}

\begin{document}

\renewcommand{\topfraction}{1.0}
\renewcommand{\bottomfraction}{1.0}
\renewcommand{\textfraction}{0.0}

\title{Relative orbit orientation in several resolved multiple systems}


\author{Andrei Tokovinin}
\affil{Cerro Tololo Inter-American Observatory, Casilla 603, La Serena, Chile}
\email{atokovinin@ctio.noao.edu}

\author{David W. Latham} 
\affil{Harvard-Smithsonian Center for Astrophysics, 60
Garden Street, Cambridge, MA 02138, USA}
\email{dlatham@cfa.harvard.edu}

\begin{abstract}
This work extends the still modest number of multiple stars with known
relative orbit orientation.  Accurate astrometry and radial velocities
are used jointly to compute or  update outer and inner orbits in three
nearby  triple systems HIP  101955 (orbital  periods 38.68  and 2.51
years), HIP  103987 (19.20  and 1.035 years),  HIP 111805  (30.13 and
1.50 years)  and in  one quadruple system  HIP 2643  (periods 70.3,
4.85 and 0.276   years), all composed of solar-type stars.  The masses
are  estimated from  the  absolute magnitudes  and  checked using  the
orbits.  The ratios of outer to  inner periods (from 14 to 20) and the
eccentricities of  the outer orbits  are moderate.  These  systems are
dynamically stable,  but not  very far from  the stability  limit.  In
three  systems all orbits  are approximately  coplanar and  have small
eccentricity,  while in  HIP 101955  the inner  orbit with  $e=0.6$ is
highly inclined.
\end{abstract}


\section{Introduction}
\label{sec:intro}

Orbits of  planets in the Solar  system, as well as  in many exoplanet
systems \citep{Fabrycky2014}, are located in one plane, presumably the
plane  of  the protoplanetary  disk.   Some  multiple stellar  systems
\citep[e.g.   HD~91962,][]{Planetary}  have  a  similar  ``planetary''
architecture and could also be formed in a disk.  However, this is not
the universal rule.  There  are triple stars with nearly perpendicular
orbits,  like  Algol, or  even  counter-rotating  triple systems  like
$\zeta$~Aqr  \citep{ZetaAqr}.   Similarly,  there  exist  non-coplanar
exoplanetary  systems  such as  $\nu$~And  \citep{McArthur2010} and  close
binaries   with   misaligned   stellar   spins   \citep{Albrecht2014}.
Dynamical interactions with other stars or planets are often evoked to
explain  the  misalignment.   In   very  tight  stellar  systems  such
interactions must  be internal (between members)  rather than external
(with  other stars  in the  cluster).   Accretion of  gas with  random
angular  momentum  during star  formation  is  another promising,  but
poorly explored mechanism of misalignment.

Orbit orientation  in triple stars  provides observational constraints
on the angular  momentum history relevant to the  formation of stellar
systems, stars,  and planets.  However, measurement  of relative orbit
orientation  in triple  stars  is challenging.   Both  orbits must  be
resolved (either  directly or astrometrically), the  sense of rotation
must  be inferred  from the  radial  velocities (RVs),  and the  outer
period must  be not  too long for  a reasonable orbit  coverage. These
conditions are met only for a small number of nearby multiple systems.
The Sixth Catalog of  Visual Binary Orbits \citep[][VB6]{VB6} contains
62 candidates of varying orbit  quality, mostly without RV data.  
  Without thorough re-assessment and  filtering, the VB6 sample is not
  suitable  for  statistical  study  of relative  orbit  orientation.
Long-baseline  stellar interferometers  help in  resolving  closer and
faster  subsystems, but require  substantial efforts,  contributing so
far only a handful of cases \citep[e.g.,][]{Kervella2013,Schaefer2016}

Motion in  a triple  system can be  described by two  Keplerian orbits
only approximately because dynamical interaction between the inner and
outer subsystems  constantly changes their orbits.  The  time scale of
this  evolution is  normally much  longer than  the time  span  of the
observations, so the orbits represent the ``instantaneous'' osculating
elements  in the  three-body  problem. Knowing  these  orbits and  the
masses,  the secular  dynamical evolution  can be  studied numerically
\citep{Xu2015}.

In  this work,  we  study  four multiple  systems  to determine  their
relative orbit orientation and period ratio as accurately as possible.
We selected  candidates with modest  period ratios and  moderate outer
eccentricity,  resembling  in  this  sense HD~91962.   Integer  period
ratios   would  suggest  potential   mean  motion   resonances.   Such
resonances   are   commonly  found   in   multi-planet  systems   
  \citep{Fabrycky2014}, but are  not documented in stellar multiples;
the case of HD~91962 with a period ratio of 18.97$\pm$0.06 remains, so
far, unique \citep[see however][]{Zhu2016}.

\begin{figure}
\epsscale{1.0}
\plotone{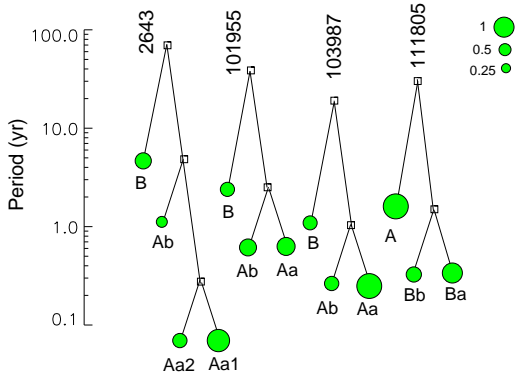}
\caption{Mobile diagrams of four multiple systems. Squares denote
 the systems (period scale on the left), green circles -- stars, with the
  circle diameter approximately proportional to the mass.  
\label{fig:mobile}  }
\end{figure}

Basic   data  on   the  four   multiple  systems   are   presented  in
Table~\ref{tab:objects};       the       mobile      diagrams       in
Figure~\ref{fig:mobile}  illustrate their  hierarchical  structure and
periods.  The range of periods is similar to that in the solar system.
The  last   two  columns   of  the  Table   give  the   parallax  from
\citet{HIP2} and  the  dynamical parallax  computed
here from  the orbital elements  and estimated masses.  HIP~2643  is a
known visual binary containing  two spectroscopic subsystems (hence it
is  quadruple); we detect  here astrometric  perturbations from  the 5
year subsystem.  The remaining three  triple stars have both inner and
outer pairs directly resolved, with their orbits already listed in the
VB6.     HIP~103987   and    111805   were    recently    studied   by
\citet[][hereafter  H15]{H15}.    We  use  the   available  astrometry
together  with the  new speckle  observations and  the RVs  to compute
combined orbits, accounting  also for the ``wobble'' in  the motion of
the  outer binary  caused by  the subsystem.

\begin{deluxetable*}{rr  c l ccc   c c }
\tabletypesize{\scriptsize}     
\tablecaption{Basic parameters of multiple systems
\label{tab:objects} }  
\tablewidth{0pt}                                   
\tablehead{                                                                     
\colhead{HIP} & 
\colhead{HD} & 
\colhead{WDS} & 
\colhead{Spectral} & 
\colhead{$V$} & 
\colhead{$B-V$} & 
\colhead{$K$} & 
\colhead{$\pi_{\rm HIP2}$} &  
\colhead{$\pi_{\rm dyn}$} \\
&   &  
\colhead{(J2000)} & 
\colhead{type} & 
\colhead{(mag)} &
\colhead{(mag)} &
\colhead{(mag)} &
\colhead{(mas)} &
\colhead{(mas)}
}
\startdata
 2643  & 2993   & 00334+4006 & F8  & 7.75 & 0.52  & 6.35 & 17.53$\pm$0.95 & 16.1 \\   
101955 & 196795 & 20396+0458 & K5V & 7.84 & 1.24  & 4.74 & 59.80$\pm$3.42 & 59.0 \\
103987 & 200580 & 21041+0300 & F9V & 7.31 & 0.54  & 5.79 & 19.27$\pm$0.99 & 23.2 \\
111805 & 214608 & 22388+4419 & F9V & 6.83 & 0.55  & 5.32 & 26.18$\pm$0.60 & 24.1
\enddata
\end{deluxetable*}

Section~\ref{sec:obs}  presents the  data used  in this  work  and the
methods common to all objects.  Then each multiple system is discussed
individually in Sections 3 to 6. The work is summarized in Section~7.

\section{Observations and their analysis}
\label{sec:obs}

\subsection{Astrometry}
\label{sec:speckle}

The  outer   subsystems  are  classical   visual  binaries.   Historic
micrometric measurements and  modern speckle interferometric data have
been obtained from  the WDS database on our  request. Additionally, we
secured new speckle astrometry  and relative photometry of two systems
at the 4.1-m SOAR telescope \citep{SOAR15}. Accurate modern astrometry
reveals  ``wobble'' in the  motion of  the outer  pairs caused  by the
subsystems, even  when those are not directly  resolved.  The 180\degr
~ambiguity of position angle in the standard speckle method is avoided
in the case of triple systems, where the orientation of the outer pair
 is known from micrometer  measures and {\it Hipparcos} and defines the
orientation of the inner pair  as well.  The observations presented in
H15  use the  image reconstruction  technique that  does not  have the
180\degr ~ambiguity.  

\subsection{Radial velocities}
\label{sec:RV}

Published RVs are  used here together with the new  data. The RVs were
measured       with      the       CfA       Digital      Speedometers
\citep{Latham1985,Latham1992},   initially  using   the   1.5-m  Wyeth
Reflector  at  the Oak  Ridge  Observatory  in  the town  of  Harvard,
Massachusetts, and  subsequently with the  1.5-m Tillinghast Reflector
at  the Whipple Observatory  on Mount  Hopkins, Arizona.   Starting in
2009  the  new fiber-fed  Tillinghast  Reflector Echelle  Spectrograph
\citep[TRES;][]{TRES}  was used.  The  spectral resolution  was 44,000
for  all three  spectrographs, but  the typical  signal-to-noise ratio
(SNR) per  resolution element of 100  for the TRES  observations was a
few times higher than for the CfA Digital Speedometer observations.

The light of all systems  except HIP~111805 is dominated by the bright
primary component.   Therefore we  followed our standard  procedure of
using one-dimensional correlations of each observed spectrum against a
synthetic template drawn from  our library of calculated spectra.  The
RV zero  point for each spectrograph was  monitored using observations
of  standard stars,  of daytime  sky, and  of minor  planets,  and the
velocities were all  adjusted to the native system  of the CfA Digital
Speedometers. To get onto the  absolute velocity system defined by our
observations of minor planets,  about 0.14 km~s$^{-1}$ should be added
to the RVs.  These velocities are  all based on correlations of just a
single echelle order centered on the Mg b triplet near 519\,nm, with a
wavelength  window of  4.5\,nm for  the CfA  Digital  Speedometers and
10.0\,nm for TRES.

Two objects,  HIP 101955  and 103987, were  observed in 2015  with the
CHIRON echelle  spectrograph \citep{CHIRON} at the 1.5  m telescope at
CTIO with a  spectral resolution of 80,000.  The  RVs were measured by
cross-correlation  of  these  spectra   with  the  digital  mask;  see
\citep{CHIRON-1} for further details.

\subsection{Orbit calculation}
\label{sec:orb}

\begin{figure}
\epsscale{0.8}
\plotone{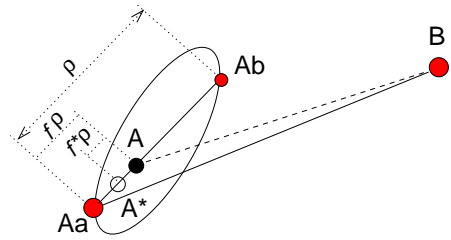}
\caption{Scheme of  a resolved triple  star with the inner  pair Aa,Ab
  and the outer  pair A,B. Red circles denote  stars, the black circle
  is the center of mass   A, the empty circle is the photo-center
  A$^*$.
\label{fig:trip} }
\end{figure}

The orbital  elements and  their errors were  determined with  the IDL
code    {\tt    orbit3.pro}\footnote{The    code    is    posted    at
  \url{http://dx.doi.org/10.5281/zenodo.321854}}       that       fits
simultaneously  the inner  and outer  orbits using  both  the resolved
measures and  the RVs.  It  describes the triple system  ``from inside
out'', as  the first inner pair  Aa,Ab and the second  outer pair A,B,
where A denotes the center of  mass of Aa,Ab.  The motion of the inner
pair depends on  the 10 inner elements. As the center  of mass A moves
in the  outer orbit, the RVs  of Aa and Ab  are sums of  the inner and
outer orbital velocities, while the RV  of B depends only on the outer
elements. For the positional  measurements, the situation is reversed:
the position  of the  inner pair depends  only on the  inner elements,
while the position of the outer pair includes the wobble term.

Figure~\ref{fig:trip} explains the wobble. The outer elements describe
the motion  of B around the center  of mass A. However,  the center of
mass is not  directly observed. Instead the measurements  of the outer
pair give the vector Aa,B if the subsystem is resolved or refer to the
photo-center A$^*$,B otherwise. The primary Aa moves around the center
of  mass  with  an  amplitude  reduced  by  the  wobble  factor  $f  =
q_1/(1+q_1)$ compared  to the inner  separation Aa,Ab, where  $q_1$ is
the inner  mass ratio.  For  the photo-center, the  appropriate wobble
factor becomes $f^* = f - r_1/(1+r_1)$, where $r_1$ is the light ratio
in the inner pair. The apparent trajectory of Aa,B or A$^*$,B includes
the wobble, it is not a closed ellipse.

The 20 orbital elements (10 inner  and 10 outer) are given as input to
the  program  and then  corrected  iteratively  to  reach the  $\chi^2$
minimum.   Errors  of  the  positional  measures  are  assumed  to  be
isotropic  (transverse   equals  radial).   The   errors  of  position
measurements and RVs are balanced when the condition $\chi^2/M \sim 1$
is reached  for each data set, where  $M$ is the number  of degrees of
freedom.  Errors of the outliers are increased to reach this balance.
The common  systemic velocity  $V_0$ is ascribed  to the  outer system
(element 20), while the wobble  factor $f$ is stored as the element number
10. Currently the code uses only one common wobble factor for all measures
of the outer pair.

 In  two objects, HIP  2643 and HIP  103987, the inner  subsystem is
  either unresolved or has  questionable measures.  The orientation of
  the inner orbit is then found  only by modeling the wobble.  In such
  cases,  the inner  semimajor axis  $a_1$ and  the wobble  factor $f$
  cannot be determined separately. We  have chosen to fix $a_1$ to its
  estimated  value, while the  wobble amplitude  is still  fitted freely
  through $f$. 

When the tertiary  component is brighter than the  inner subsystem (it
is  usually  denoted  then  as   A),  it  is  still  considered  as  a
``tertiary''  by the code.   In such  case, the  wobble factor  $f$ is
negative and the outer  elements $\Omega_2$ and $\omega_2$ are flipped
by $180^\circ$.

The   orbital    elements   and    their   errors   are    listed   in
Table~\ref{tab:orb}. Its first
column identifies each subsystem by the {\it Hipparcos} number and, in
the  following   line,  by  the  ``discoverer   code''  and  component
designations  joined by  the  comma. The  following  columns give  the
period $P$, the  epoch of periastron $T_0$, the  eccentricity $e$, the
semimajor  axis  $a$,  the   position  angle  of  the  ascending  node
$\Omega_{\rm A}$ (for the epoch  J2000) and the argument of periastron
$\omega_{\rm  A}$ (both angles  refer to  the primary  component), the
orbital inclination $i$,  the RV amplitudes $K_1$ and  $K_2$. The last
column contains  the systemic velocity  $V_0$ for the outer  orbit and
the wobble factor $f$ for the inner orbit.

Table~\ref{tab:pos},  available in full electronically, lists the
positional  measures  and  their  residuals.  Its  first  two  columns
identify  the  pair by  its  {\it  Hipparcos}  number and  the  system
designation. The following columns contain (3) the date of observation
in  Besselian  years,  (4)   the  position  angle  $\theta$,  (5)  the
separation $\rho$, (6) the assumed error $\sigma$, (7) residual to the
orbit in  angle and (8) in  separation. The last  column (9) indicates
the measurement  technique, as  described in the  notes to  the Table.
Table~\ref{tab:rv},    also  available  in  full  electronically,
contains the  RVs. Its first  two columns specify the  {\it Hipparcos}
number and the component. Then follow (3) the Julian date, (4) the RV,
(5) its  error, and (6)  the residual. The  last column (7)  gives the
source of the RV, as explained in the notes.

\begin{deluxetable*}{l l rrr rrr r ccc}
\tabletypesize{\scriptsize}
\tablewidth{0pt}
\tablecaption{Orbital Elements \label{tab:orb}}
\tablehead{
\colhead{HIP/system} &
\colhead{$P$} & 
\colhead{$T_0$} &
\colhead{$e$} & 
\colhead{$a$} & 
\colhead{$\Omega_{\rm A}$} &
\colhead{$\omega_{\rm A}$} &
\colhead{$i$}  &
\colhead{$K_1$}  &
\colhead{$K_2$}  &
\colhead{$V_0$, $f$}  \\
\colhead{Other designation} &   
\colhead{(yr)} & 
\colhead{(yr)} &
\colhead{ } & 
\colhead{($''$)} & 
\colhead{(\degr)} &
\colhead{(\degr)} &
\colhead{(\degr)} &
\colhead{(km~s$^{-1}$)} &
\colhead{(km~s$^{-1}$)} &
\colhead{(km~s$^{-1}$)} &
}
\startdata
2643/outer    & 70.34        & 1983.62    & 0.331       & 0.393      & 118.9    & 137.8   &112.3 & (3.2)      & (6.9)       & $-$1.37\\
~~HO 3 A,B    &  $\pm$1.36   & $\pm$0.60 & $\pm$0.032  & $\pm$0.020  & $\pm$0.5 & $\pm$2.7 &$\pm$0.9     & \ldots &  \ldots  &  $\pm$0.19 \\
2643/middle   & 4.849       & 1994.927    & 0.138      & (0.058)  & 100.7  & 132.3    & 94.0           & 4.857      & \ldots   & 0.217    \\
Aa,Ab         & $\pm$0.020   & $\pm$0.010   & $\pm$0.028 & fixed & $\pm$7.9  & $\pm$7.67&  $\pm$ 8.2    & $\pm$0.12 & \ldots &  $\pm$0.034     \\
2643/inner    & 0.27595      & 1997.1620   & 0.1986     & \ldots    & \ldots    & 113.4    & \ldots   & 12.493      &  \ldots      &  (0.0)    \\
Aa1,Aa2       & $\pm$0.00002   & $\pm$0.0012  & $\pm$0.0073  & \ldots     & \ldots & $\pm$1.7 &  \ldots    & $\pm$0.109 & \ldots    & \ldots         \\
101955/outer  & 38.6790        & 2016.110   & 0.118     & 0.855      & 127.6   & 233.4   & 87.40   & 2.66      & \ldots   & $-$41.11   \\
~~KUI 99 A,B  &  $\pm$0.031    & $\pm$1.32 & $\pm$0.016 & $\pm$0.110 &$\pm$0.08& $\pm$0.5&$\pm$0.05& $\pm$0.40 &  \ldots       &  $\pm$0.08 \\
101955/iner   & 2.51013        & 2000.518  & 0.6170      & 0.1242      &147.1       & 109.7   & 24.1    & 3.27       & 6.93   & 0.457  \\
~~BAG 14 Aa,Ab&  $\pm$0.00052  & $\pm$0.004 & $\pm$0.0047& $\pm$0.0011 &$\pm$1.8    & $\pm$1.8 & $\pm$1.7 &$\pm$0.12 &$\pm$0.71& $\pm$0.005  \\
103987/outer  & 19.205         & 2006.259   & 0.1743      & 0.2195      & 102.8    & 17.6     & 65.1    & 4.005      & 9.58        & $-$1.97    \\
~~WSI 6 A,B   &  $\pm$0.080    & $\pm$3.60  & $\pm$0.0083 & $\pm$0.0013 & $\pm$0.5 & $\pm$2.6 &$\pm$1.0 & $\pm$0.082 &  $\pm$0.22  &  $\pm$0.05 \\
103987/inner  & 1.03483        & 2014.6223   & 0.0934     &  0.0284     & 97.3     & 124.9    & 68.6    & 9.528      & \ldots      & 0.350  \\
~~DSG 6 Aa,Ab &  $\pm$0.00008    & $\pm$0.0089 & $\pm$0.0040& fixed     & $\pm$12.5& $\pm$3.1 &$\pm$13.7& $\pm$0.058 & \ldots      & $\pm$0.062 \\
111805/outer  & 30.127         & 2010.179   & 0.324      & 0.3361      & 154.25   & 84.92    & 88.28   & 6.06      & 8.60     & $-$22.58   \\
~~HDO 295 B,A &  $\pm$0.031    & $\pm$0.073 & $\pm$0.004 & $\pm$0.0015 & $\pm$0.09& $\pm$0.18&$\pm$0.10& $\pm$0.14 & $\pm$0.23&$\pm$0.08 \\
111805/iner   & 1.5012         & 1986.093   & 0.022     & 0.0385      & 334.5    & 232.9    & 85.80   & 13.13     & 19.21    & $-$0.330      \\
~~BAG 15 Ba,Bb&  $\pm$0.0004   & $\pm$0.093 & $\pm$0.011 & $\pm$0.0010 & $\pm$1.0 & $\pm$22.3&$\pm$1.6 & $\pm$0.25 & $\pm$3.1 &  $\pm$0.015
\enddata
\end{deluxetable*}

\begin{deluxetable*}{ll   c  ccc   cc l }
\tabletypesize{\scriptsize}     
\tablecaption{Relative positions and residuals (fragment)
\label{tab:pos} }  
\tablewidth{0pt}                                   
\tablehead{                                                                     
\colhead{HIP} & 
\colhead{Sys} & 
\colhead{Date} & 
\colhead{$\theta$} & 
\colhead{$\rho$} & 
\colhead{$\sigma$} & 
\colhead{O$-$C$_\theta$} & 
\colhead{O$-$C$_\rho$} & 
\colhead{Ref\tablenotemark{a}} \\
&   &  
\colhead{(year)} & 
\colhead{(\degr)} & 
\colhead{(\arcsec)} &
\colhead{(\arcsec)} &
\colhead{(\degr)} &
\colhead{(\arcsec)} &
}
\startdata
  2643 & A,B &  1885.8100 &    121.2 &   0.5000 &   0.1000 &     -6.4 &   0.0055 & M  \\
  2643 & A,B &  1948.7900 &    137.8 &   0.3400 &   0.1000 &      0.6 &  -0.0823 & M \\
  2643 & A,B &  1954.9800 &    130.1 &   0.4900 &   0.1000 &      1.5 &   0.0036 & M   
\enddata
\tablenotetext{a}{
G: DSSI at Gemini-N;
H: {\it Hipparcos};
M: micrometer measures; 
S: speckle interferometry at SOAR;
s: other speckle interferometry
}
\end{deluxetable*}

\begin{deluxetable*}{ll   c  ccc   l }
\tabletypesize{\scriptsize}     
\tablecaption{Radial velocity and residuals (fragment)
\label{tab:rv} }  
\tablewidth{0pt}                                   
\tablehead{                                                                     
\colhead{HIP} & 
\colhead{Comp} & 
\colhead{JD} & 
\colhead{RV} & 
\colhead{$\sigma_{\rm RV}$} & 
\colhead{O$-$C} & 
\colhead{Ref\tablenotemark{a}} \\
&   &  
\colhead{+2400000} & 
\multicolumn{3}{c}{ (km\,s$^{-1}$) } &
}
\startdata
  2643 & Aa1 & 48851.5080 &      6.940 &      0.490 &     -0.826 & T \\
  2643 & Aa1 & 48947.3570 &      5.330 &      0.500 &     -0.884 & L \\
  2643 & Aa1 & 49952.5990 &     -4.870 &      0.590 &     -0.703 & L \\
  2643 & Aa1 & 48913.5800 &    -12.890 &      1.700 &     -1.497 & L
\enddata
\tablenotetext{a}{
C: CHIRON;
D: D87; 
L: CfA;
L-: Cfa $-$1~km~s$^{-1}$;
T: \citet{TS02}
}
\end{deluxetable*}


\subsection{Photometry and masses}
\label{sec:models}

\begin{deluxetable}{l l cc}
\tabletypesize{\scriptsize}
\tablewidth{0pt}
\tablecaption{Magnitudes and masses \label{tab:ptm}}
\tablehead{
\colhead{HIP} &
\colhead{Comp.} & 
\colhead{$V$} & 
\colhead{${\cal M}$} \\
 & & 
\colhead{(mag)} & 
\colhead{(${\cal M}_\odot$)} 
}
\startdata
2643 & Aa1 & 7.97 & 1.22 \\
     & Aa2 & (14.9) & (0.36) \\
     & Ab  & (14.5) & 0.42 \\
     & B   & 9.60   & 0.91 \\ 
101955 & Aa & 8.39 & 0.74 \\ 
       & Ab & 9.74 & 0.62 \\ 
       & B  & 9.47 & 0.65 \\ 
103987 & Aa & 7.46 & 1.15 \\
       & Ab & (12.5) & 0.56 \\
       & B  & 9.62   &  0.67 \\
111805 & A  & 7.48   & 1.14 \\
       & Ba & 7.98   & 1.03 \\
       & Bb & 9.25   & 0.85 
\enddata
\end{deluxetable}

The relative photometry  of the resolved pairs is  available from {\it
  Hipparcos} and speckle  interferometry.  This defines the individual
magnitudes of the components and, knowing the distance, their absolute
magnitudes.  All  components are normal main  sequence stars, allowing
us to estimate  their masses from the standard  relations. We use here
the polynomial  approximation of the absolute  magnitude dependence on
mass and wavelength from  \citep{FG14}.  The magnitudes, distance, and
masses constitute the model of each object (Table~\ref{tab:ptm}). 
  Magnitudes not measured directly are given in brackets.

The sum  of the estimated  masses does not  always match the  mass sum
computed from  the orbital elements and the  {\it Hipparcos} parallax.
The  latter can be  biased  by  the complex  orbital  motion in  multiple
systems  that  has  not  been  accounted  for  in  the  original  {\it
  Hipparcos}  data reduction  \citep{Soderhjelm1999}.   Therefore, the
distances  used here  are derived  from the  mass sum  and  the orbits
(so-called     dynamical    parallaxes     $\pi_{\rm     dyn}$,    see
Table~\ref{tab:objects}).  The RV amplitudes are used as a check, with
the  caveat  of  a  potential  RV  bias due  to  blending  with  other
components.  The wobble amplitude and the combined color of the system
are additional  ways to  check the consistency  of the  system models.
For all multiple systems studied here, the minor discrepancies between
various estimates  of masses  and magnitudes can  be explained  by the
errors and biases.

\section{HIP 2643}
\label{sec:2643}

\begin{figure}
\epsscale{1.0}
\plotone{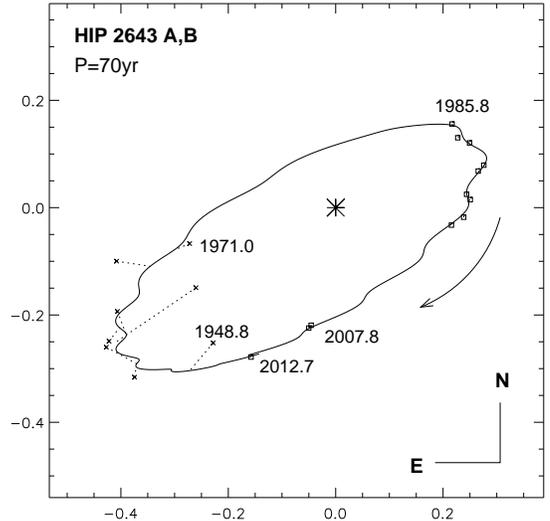}
\caption{Orbit of HIP 2643 A,B in  the sky. The primary component A is
  located  at  the  coordinate  origin  (asterisk); the  scale  is  in
  arcseconds.  Squares  and crosses  denote relative positions  of the
  secondary  component  B   measured  by  speckle  interferometry  and
  micrometers,  respectively;  the  short  dotted  lines  connect  the
  measurements  to the calculated  positions on  the orbit.   The wavy
  line is one complete orbit of the outer pair with the wobble.
\label{fig:2643}  }
\end{figure}

\begin{figure*}
\epsscale{1.0}
\plotone{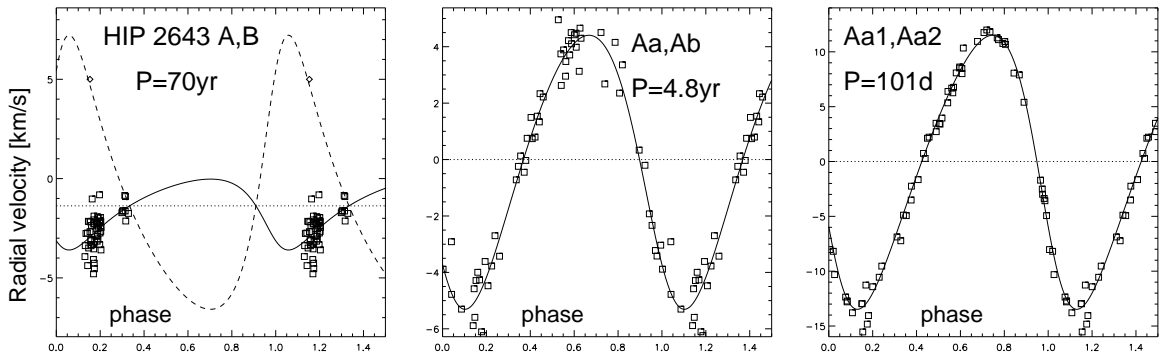}
\caption{RV  curves of HIP  2643.  Left:  outer orbit,  middle: middle
  orbit,  right: inner  orbit. The  full and  dashed lines  denote the
  curves for  the primary and  secondary components.  The  squares are
  the measured RVs where the motion in other orbits is subtracted. The
  diamond in  the left-hand plot  is a fake  RV of the component  B to
  illustrate the blending.
\label{fig:2643RV}  }
\end{figure*}

HIP 2643  (HD 2993) is known as  a visual binary HO~3  or ADS~463.  It
has been first  resolved in 1887 by Holden, so  the measures cover 1.8
outer  periods.  The  visual orbits  of A,B  were computed  by several
authors;  the latest  orbit by  \citet{Msn2014a} has  $P=69.15$ years.
Independently, D.~L.  has discovered the RV variations with periods of
100  and  1485   days,  meaning  that  the  primary   component  is  a
spectroscopic triple, although  only one star is seen  in the spectra.
The whole system is therefore a 3+1 quadruple with a 3-tier hierarchy,
like HD 91962.  The innermost 101-day pair is Aa1,Aa2, the middle pair
is Aa,Ab, and the outer visual pair is A,B (Figure~\ref{fig:mobile}). 

Accurate  speckle measures  of A,B  available since  1985 allow  us to
detect  the ``wobble''  caused by  the middle  subsystem Aa,Ab  and to
determine     the     elements     of    its     astrometric     orbit
(Figure~\ref{fig:2643}). As our code cannot deal with quadruple stars,
we first  fitted the  two inner spectroscopic  orbits (the  latest RVs
were reduced by 1\,km~s$^{-1}$ to account for the outer orbit).  Then
the RV variations  caused by the 101 day  inner orbit were subtracted,
and  the   corrected  RVs  were  used  jointly   with  the  positional
measurements  to fit  the middle  and outer  orbits. As  there  are no
resolved  measures of  the middle  subsystem, its  semimajor  axis was
estimated from the  masses and period and fixed  to 58\,mas, while the
wobble factor $f$ was fitted.

Although only  two stars  are directly observed,  we can  evaluate the
masses  of all four  components.  To  begin with,  we assume  that the
innermost  orbit has  a large  inclination (this  is justified  in the
following   paragraph).\footnote{    We  do   not   determine  the
  orientation of  the inner orbit in  this paper.} Then  the inner RV
amplitude and the estimated mass of  Aa1 lead to the mass of Aa2, 0.36
${\cal M}_\odot$.  The inclination of the middle orbit is known, hence
the  mass of  Ab is  0.42 ${\cal  M}_\odot$, while  the mass  of  B is
estimated from  its absolute  magnitude.  The outer  mass sum  of 2.91
${\cal  M}_\odot$ leads to  the dynamical  parallax of  16.1\,mas; the
{\it Hipparcos} parallax of 17.5\,mas is likely biased.

Taking the masses listed  in Table~\ref{tab:ptm}, we convert them back
to the $V$  and $K$ magnitudes using the  same standard relations.  It
turns out  that the  spectroscopic secondaries Aa2  and Ab  are indeed
faint and do  not influence the combined photometry.   The modeled and
measured combined $K$ magnitudes  are 6.26 and 6.35 mag, respectively,
so the model  reproduces the actual $V-K$ color  reasonably well.  The
wobble factor $f=0.22$  leads to the mass ratio of  0.28 in the middle
orbit, while the  adopted masses imply $q =  0.27$, in good agreement.
If the innermost orbit had a  small inclination, the mass of Aa2 would
be larger, and the wobble factor would be smaller than measured.

Given the estimated masses, we evaluate the RV amplitudes in the outer
orbit, $K_1 = 3.2$ km~s$^{-1}$  and $K_2 = 6.9$ km~s$^{-1}$.  The free
adjustment leads the much smaller  $K_1 = 1.4$ km~s$^{-1}$.  The RV of
B  in  Figure~\ref{fig:2643RV} is  a  fake  point  added to  show  the
expected  RV curve  for  the  visual secondary  that  is not  actually
measured.  Blending with the lines  of B likely explains the too small
RV amplitude derived  by the free fit to the RVs  and increases the RV
errors of  Aa1, compared  to a truly  single star (rms  residuals 0.56
km~s$^{-1}$).  The  RV residuals indeed correlate  positively with the
RV, as expected from the blending effect.  The sign of the RV trend in
the  outer orbit establishes  its correct  node.  New  RV measurements
would be helpful for a better definition of the outer orbit and of all
the periods.  The spectrum of B should be detectable as it contributes
0.21 fraction of the combined light in the $V$ band.

The inner period  ratio is 17.57$\pm$0.07, the outer  period ratio is
14.43$\pm$0.28.  The angle $\Phi$  between the middle and outer orbits
is  $25\fdg4 \pm  8\fdg5$. With  such relative  inclination, the
orbit  of Aa,Ab  precesses, but  does not  experience  the Kozai-Lidov
cycles.  The small eccentricity  of all orbits supports indirectly the
absence of such cycles and the approximate coplanarity of all orbits.

\section{HIP 101955}
\label{sec:101955}

\begin{figure}
\epsscale{1.0}
\plotone{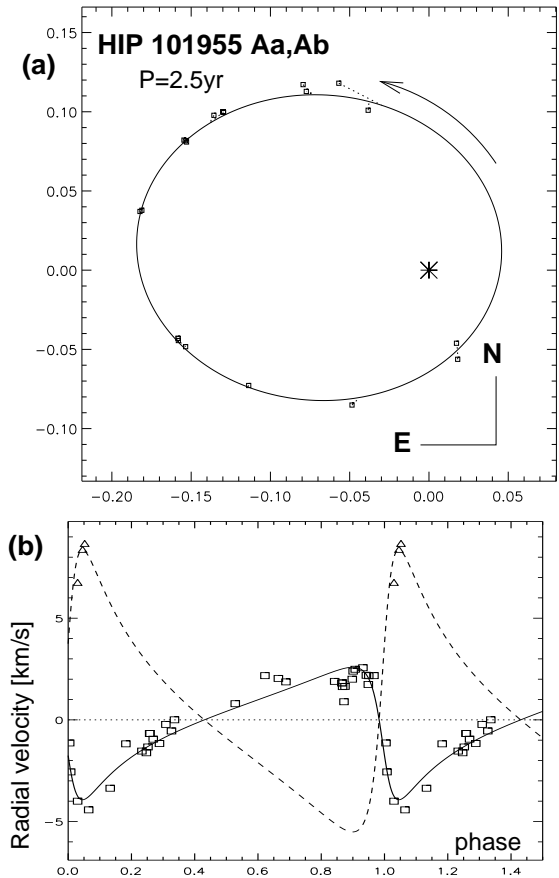}
\caption{The inner orbit of HIP 101955 Aa,Ab. Top (a): orbit in the plane of the
  sky, bottom (b): the RV curve with the outer
  orbit subtracted. 
\label{fig:101955in}  }
\end{figure}

\begin{figure}
\epsscale{1.0}
\plotone{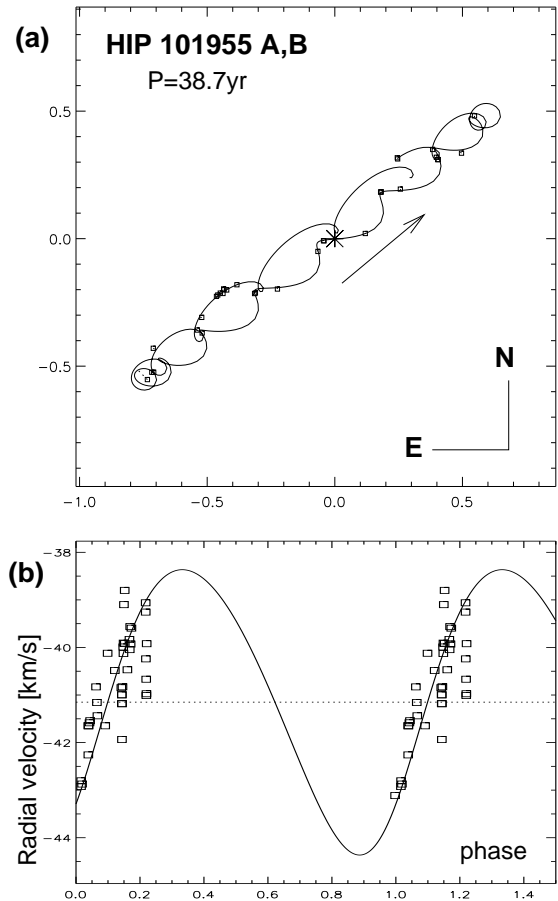}
\caption{Outer orbit of HIP 101955 A,B.  Top (a): orbit in the plane
  of  the sky;  only accurate  speckle  measures are  plotted for  one
  complete outer period.  Bottom (b): the RV curve. 
\label{fig:101955out}  }
\end{figure}

This is a  nearby (17\,pc) triple system HD~196795  or GJ~795.  It has
an extensive literature.  The  inner subsystem Aa,Ab was discovered by
\citet{D87} (hereafter  D87) using CORAVEL and later  resolved for the
first  time by \citet{Balega2002};  \citet{Malogolovets2007} presented
detailed study  of this triple system.  Unlike  other systems featured
here,  the inner  orbit has  a  substantial eccentricity  and the  two
orbits have large  mutual inclination.   Figure~\ref{fig:101955in}
  displays  the inner  orbit, while  Figure~\ref{fig:101955out} shows
the trajectory of  A,B with the wobble.  Only  the speckle measures of
A,B since 1981 are used to  fit the two orbits jointly, with the outer
period  fixed   to  its  value found  by  using  all data.   The
semimajor axis  of the wobble is 55.1$\pm$0.6\,mas.   The weighted rms
residuals for  both A,B and  Aa,Ab are $\sim$3\,mas in  position, 0.48
and 0.58 km~s$^{-1}$ for the RVs of Aa and Ab, respectively.

One spectrum  of HIP~101955 has been  taken with CHIRON in  2015 on JD
2457261.  Its cross-correlation function (CCF) with the binary mask is
an asymmetric blend that can  be fitted by two Gaussians.  The fainter
component is in fact  a blend of Ab and B, as  their RVs were close at
that time.  The  ratio of the dip areas  of Ab+B and Aa in  the CCF is
0.50,  or  $\Delta  m  =  0.75$ mag,  matching  roughly  the  resolved
photometry (the  system model predicts  0.45 mag).  The rms  widths of
the  CCF  dips  of  Aa   and  Ab+B  are  4.12  and  5.11  km~s$^{-1}$,
respectively.

The RVs measured by Duquennoy are likely affected by blending (CORAVEL
did  not resolve  the  blends,  except on  two  occasions).  Owing to  this,
the  ``spectroscopic'' masses  of Aa and  Ab derived  from the
combined inner orbit   are too small (0.7 and 0.4 ${\cal M}_\odot$).

The  system  model  starts   with  the  combined  $V=7.84$,  the  {\it
  Hipparcos}  parallax, and the  magnitude differences  $\Delta V_{\rm
  Aa,Ab} =  1.35$ mag \citep{Malogolovets2007}, $\Delta  V_{\rm A,B} =
1.35$ mag ({\it Hipparcos}).   Individual magnitudes of the components
and their estimated masses  are listed in Table~\ref{tab:ptm}, leading
to the mass sum of of 2.02  ${\cal M}_\odot$ for AB.  The orbit of A,B
matches  this mass  sum  for  a parallax  of  59.0\,mas, in  excellent
agreement  with 58.8\,mas  determined by  \citet{Soderhjelm1999}.  The
mass sum derived  from the inner orbit is  then 1.49 ${\cal M}_\odot$,
while the model  gives 1.36 ${\cal M}_\odot$.  The  model predicts the
combined $K$ magnitude of 4.62 mag, the observed one is 4.74 mag.

The  adopted masses imply  $q_{\rm Aa,Ab}=0.84$  and match  the wobble
amplitude  that corresponds to  $q_{\rm Aa,Ab}  = f/(1  - f)  = 0.84$.
However, the spectroscopic  inner mass ratio is 0.45;  it is biased by
the reduced RV  amplitude of Aa and strongly  contradicts the relative
photometry of  Aa,Ab.  Even  if the RV  amplitudes in the  inner orbit
were   measured  reliably,   its  small   inclination   prevents  good
independent measurement of stellar masses and distance.  The RVs of Aa
also do  not fit well the outer  orbit due to blending  with the other
components Ab and B.  The mass  sum in the outer system corresponds to
$K_1 + K_2  = 11.2$ km~s$^{-1}$ (this is a  robust estimate, given the
high  inclination), and  the mass  ratio  $q_{\rm A,B}$  leads to  the
estimated amplitudes  in the outer orbit  $K_1 = 3.0$ and  $K_2 = 8.2$
km~s$^{-1}$.  The fitted  $K_1$ in  the outer  orbit converges  to 2.66
km~s$^{-1}$.

The  period ratio  is 15.41$\pm$0.13,  the angle  between  the orbital
angular   momenta  is  $\Phi   =  64\fdg8  \pm1\fdg4$.   Strong
interaction  between the  orbits and  Kozai-Lidov cycles  are expected
\citep{Xu2015}.   \citet{Malogolovets2007}  estimated  the  period  of
these cycles as 560 years and noted that they may be observable.

\section{HIP 103987}
\label{sec:103987}

\begin{figure}
\epsscale{1.0}
\plotone{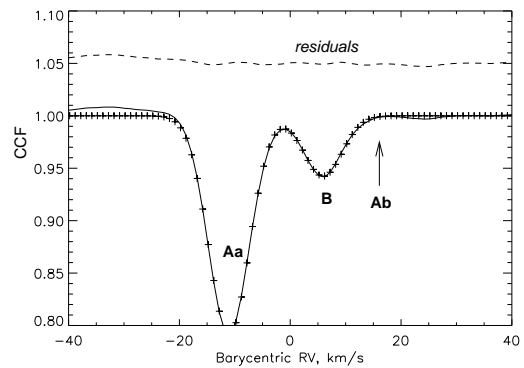}
\caption{Cross-correlation function (CCF)  of HIP~103987  on JD 2457301
  and its  approximation with  two Gaussians (crosses).  The residuals
  are  plotted in dashed  line shifted  to 1.05.  The arrow  marks the
  expected RV of the undetected component Ab. 
\label{fig:CCF}  }
\end{figure}

\begin{figure}
\epsscale{1.0}
\plotone{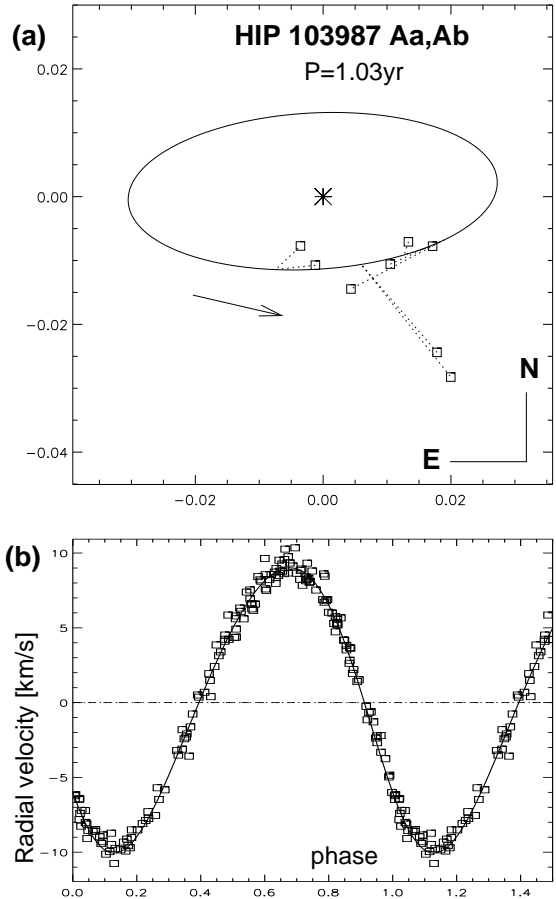}
\caption{Orbit of HIP 103987 Aa,Ab in the plane of the sky (top, a)
  and the corresponding RV curve (bottom, b).
\label{fig:103987in}  }
\end{figure}

\begin{figure}
\epsscale{1.0}
\plotone{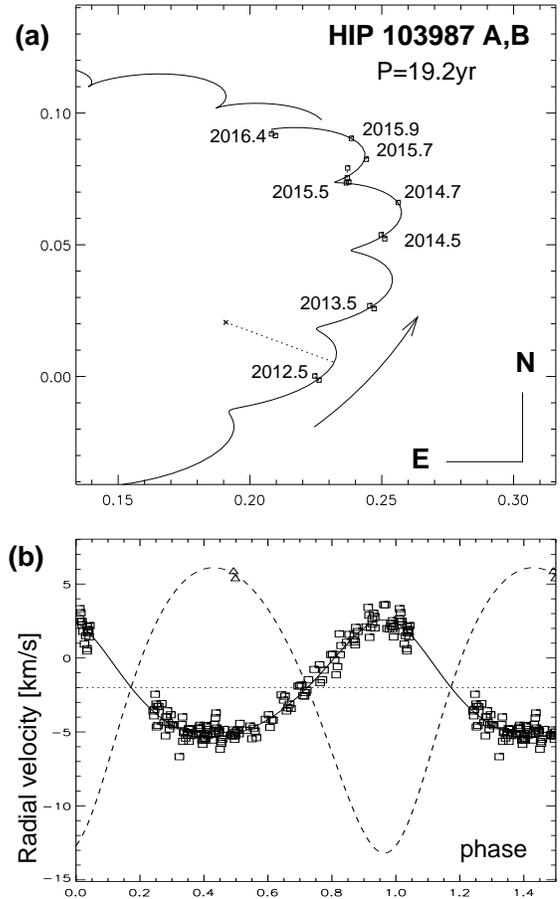}
\caption{Orbit of HIP  103987 A,B (WSI~6) in the plane  of the sky (a),
  where only a fragment of the full orbit is shown to highlight the wobble,
  and  the RV curve  (b), with  squares for  RV(Aa) and  triangles for
  RV(B).
\label{fig:103987out}  }
\end{figure}

HD~200580, alias G25-15, is a metal-poor ([Fe/H] $\sim -0.6$) multiple
system  with  a  fast  proper  motion  of  0\farcs46  per  year.   The
single-lined spectroscopic orbit with one year period was published by
\citet{Latham2002}, while  the outer system A,B was  first resolved by
\citet{Mason2001}  in 1999  and is  known as  WSI~6.   The astrometric
orbit  of the  inner subsystem  was published  by \citet{Jancart2005}.
The  inner pair Aa,Ab  was resolved  at Gemini,  and its  first visual
orbit was  published in  H15.  The available  RVs now cover  1.5 outer
periods and lead  to the spectroscopic orbits of  both inner and outer
subsystems.  The 19 year outer  period found from the RVs matches well
the visual orbit that covers  almost the full ellipse; the preliminary
21 year orbit  of A,B was published by \citet{RAO}.   The pair A,B was
observed at SOAR several times, but the inner subsystem has never been
resolved.

In 2015, the star was observed twice with CHIRON in order to get fresh
RVs and to  detect the lines of other components.   Indeed, the CCF of
the   spectrum  and  mask   is  double   (Figure~\ref{fig:CCF}).   Its
components  correspond  to  the  visual  primary  Aa  and  the  visual
secondary  B.  There  is  no trace  of  Ab, which  should  have RV  of
+15\,km~s$^{-1}$  at  the  moment  of observation;  the  non-detection
implies that Ab is at least $\sim$4 mag fainter than B and contradicts
the  speckle photometry in  H15.  Both  CCF dips  are very  narrow and
correspond to the axial rotation $V \sin i$ of 2.2 and 1.5 km~s$^{-1}$
for Aa and B, respectively.  The ratio of the CCF areas corresponds to
$\Delta m_{\rm Aa,B} = 1.43$  mag.  At SOAR we measured $\Delta y_{\rm
  A,B}  =  2.17$   mag  with  the  rms  scatter   of  0.05  mag.   The
spectroscopic $\Delta m  $ is  underestimated because  B has a
lower effective temperature and stronger lines than Aa.

The speckle measures  of A,B are accurate enough  to detect the wobble
caused  by  the subsystem  Aa,Ab  and  to  determine all  its  orbital
elements    except   $a_1$.   Figures   ~\ref{fig:103987in}  and
~\ref{fig:103987out} show  the inner  and outer orbits.   The weighted
rms residuals to  the measures of A,B in both  coordinates are 1.3 and
2.5 mas, the wobble amplitude is 9.9$\pm$1.8\,mas.  The astrometry has
adequate phase  coverage of the  inner period mainly because  the pair
has been  extensively monitored at SOAR  during 2015 with  the goal to
resolve  the  Aa,Ab at  quadratures,  where  the predicted  separation
reaches 20\,mas.  Most other measures  of A,B are from Gemini and have
an excellent  accuracy of $\sim$1\,mas.  They were  obtained at nearly
the same  phase of  the inner  orbit, as dictated  by the  Gemini time
allocations.  Joint  analysis of the  RVs and astrometry leads  to the
reliable inner orbit. The  largest correlations are +0.5 between $i_1$
and $f$  (which defines the astrometric amplitude)  and $-0.5$ between
$i_1$  and $\Omega_1$.   The wobble  is detected  at the  $5.6 \sigma$
significance level.  To test the robustness of the relative orbit
  orientation,  we fixed  $f$ to  values  0.29 and  0.41 (within  $\pm
  \sigma$ of  the nominal) and repeated  the fits.  In  both cases the
  angle $\Phi$ increased by 2\degr, less than its error. 

The subsystem Aa,Ab was resolved  at Gemini in four seasons (including
the  preliminary   data  of  2015  communicated   by  E.~Horch).   The
separation was  close to  the diffraction limit  of Gemini,  hence the
estimates of  the separation and  $\Delta m$ are  correlated. Probably
for  this reason, the  $\Delta m_{\rm  Aa,Ab} =  1.54$ mag  at 692\,nm
published  in H15  appears strongly  under-estimated  (E.~Horch, 2016,
private  communication)   and  contradicts   both  the  lack   of  the
spectroscopic detection of Ab  and its mass evaluated below. According
to Figure~1  in H15 and  $\Delta m_{\rm Aa,Ab}  = 4.1$ mag  at 692\,nm
estimated here,  the subsystem should be undetectable.   In all Gemini
runs, the pair Aa,Ab was  oriented in the North-South direction, where
the atmospheric dispersion, which is not physically compensated in the
DSSI speckle  camera, could distort the measures.   The prograde orbit
of Aa,Ab  computed in H15 from  the resolved measures  has a near-zero
inclination, which  contradicts the non-zero RV amplitude  of Aa. This
said,   the  positions   of  the   inner  pair   measured   at  Gemini
(Figure~\ref{fig:103987in}) roughly  match its orbit,  except the 2014
measures  with nearly  double  separation.  We  question the  resolved
measures of  the inner pair and  use them in the  combined orbital fit
with very small weights.

The astrometric orbit of Aa,Ab by \citet{Jancart2005} has the expected
semimajor axis of 10.4\,mas,  but corresponds to the retrograde motion
($i=162^\circ$) and has a very different $\Omega_{\rm A} = 15.6^\circ$
compared to our  orbit.  These authors have not  revised the parallax,
which is  necessary for  a one year  binary.  Moreover,  their results
could  be  biased  by  the  visual  component  B  unresolved  by  {\it
  Hipparcos}.  Therefore, this astrometric orbit should be ignored.

Owing to  the RV(B)  measured with CHIRON,  the combined orbit  of A,B
yields the  orbital parallax of 23.4$\pm$0.3\,mas  and the components'
masses 1.62$\pm$0.09  ${\cal M}_\odot$ for A  and 0.67$\pm$0.06 ${\cal
  M}_\odot$ for B. However, the orbital masses are proportional to the
cube of the  RV amplitudes, and if the  amplitudes are slightly reduced
by the  line blending  with other components,  the orbital  masses are
under-estimated.   The  {\it   Hipparcos}  parallax  of  19.3\,mas  is
evidently biased by the 1-year wobble.

We adopt the  masses of 1.15, 0.56, and 0.67  ${\cal M}_\odot$ for Aa,
Ab, and B,  respectively, based on the system  model.  They correspond
to  the  mass sum  of  2.38  ${\cal  M}_\odot$, slightly  larger  than
measured,  and  the  dynamical  parallax of  23.2\,mas,  matching  the
orbital parallax  within its error.  The inner  semimajor axis $a_{\rm
  Aa,Ab} = 28.4$\,mas  is then computed and fixed   (remember that
  the resolved  measures of Aa,Ab  are questionable, while  the wobble
  amplitude is determined by the fitted factor $f$). 

The  inner mass  ratio $q_{\rm  Aa,Ab}  = 0.49$  matches the  measured
wobble amplitude.   The spectroscopic mass of Ab calculated  from the
inner RV amplitude and inclination is 0.49 ${\cal M}_\odot$. Agreement
with the  adopted mass  of Ab would  be reached  by a 4\%  increase of
$K_1$  or for  the  inner  inclination of  $55^\circ$  instead of  the
measured $69^\circ \pm  14^\circ$.  If Ab is a  normal M-type dwarf, we
expect $\Delta m_{\rm  Aa,Ab} = 4.1$ mag at  692\,nm, much larger than
$\Delta m_{\rm Aa,Ab} = 1.5$ mag measured at Gemini and in agreement
with the spectroscopic non-detection of Ab with CHIRON. Alternatively,
Ab could be a white dwarf. 

The components' magnitudes  listed in Table~\ref{tab:ptm} are computed
by using $\Delta V_{\rm AB} =  2.13$ mag measured by speckle interferometry at
SOAR and by  further assuming that $\Delta V_{\rm Aa,Ab} =  5$ mag to match
the  adopted  mass of  Ab.   The  model  reproduces the  combined  $K$
magnitude and predicts $V  - K = 1.63$ mag; the actual  $V - K = 1.52$
mag is slightly bluer, as it should be for a low-metallicity star.
The  outer and  inner  orbits are  inclined  by $\Phi  = 6^\circ  \pm
9^\circ$, i.e. are almost coplanar.  The period ratio is 18.55$\pm$0.08.

\section{HIP 111805}
\label{sec:111805}

\begin{figure}
\epsscale{1.0}
\plotone{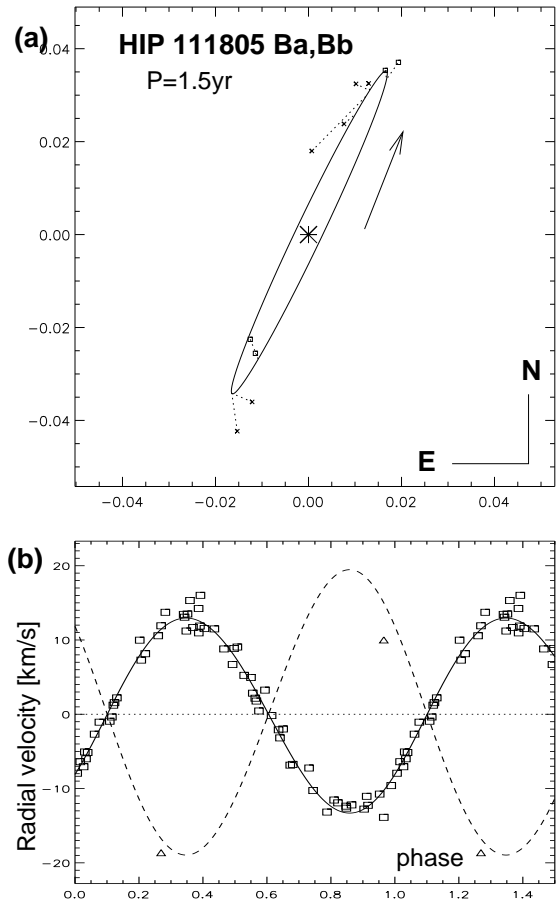}
\caption{The orbit of HIP 111805 Ba,Bb (BAG 15). Top (a): orbit in
  the sky; the measures at Gemini are plotted as squares, the
  remaining measures as crosses.  Bottom (b): the RV curve (squares
  for Ba and triangles for Bb).
\label{fig:111805in}  }
\end{figure}

\begin{figure}
\epsscale{1.0}
\plotone{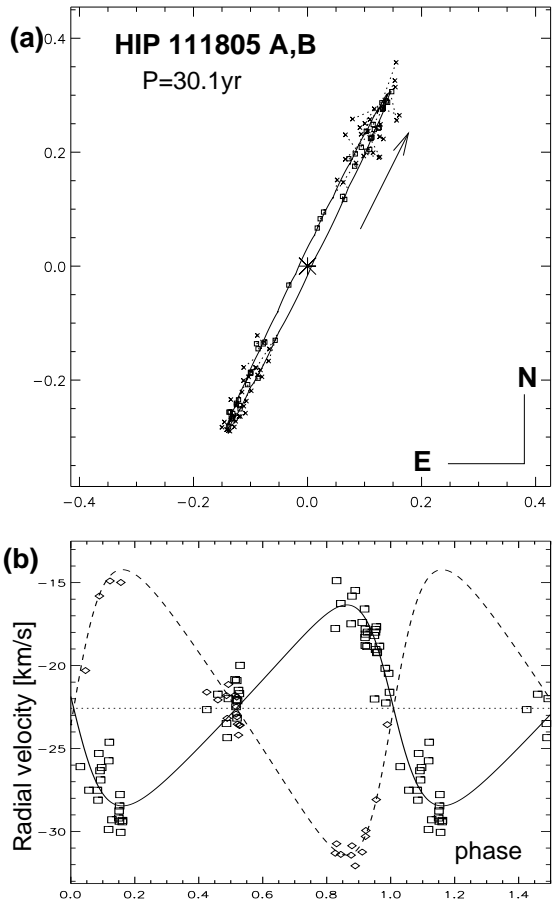}
\caption{The orbit of HIP 111805 A,B  (HDO 295). Top (a): orbit in the
  sky, bottom (b): the RV curve (squares for Ba and diamonds for A).
\label{fig:111805out}  }
\end{figure}

This system,  HD~214608, is  also metal-poor; it  was studied  in H15.
The subsystem Ba,Bb, first  resolved by \citet{Balega2002}, belongs to
the secondary  component of the well-studied visual  pair HDO~295 (ADS
16138) known since 1887. Both orbits are seen nearly edge-on  and,
 as shown below,  are well aligned, $\Phi =  2\fdg5 \pm 1\fdg5$. For
this reason, the  wobble is not obvious in the  outer orbit plot. 
  Most  speckle observations did  not resolve  the inner  pair Ba,Bb,
  which has fewer speckle measures compared to A,B.

The  spectroscopic  orbits  of  both  subsystems  were  determined  by
\citet{D87}.  Additional  RVs obtained by  D.~L.  are used  here. They
were derived by TODCOR and refer to the two brightest components A and
Ba. The  components often blend, therefore several  highly deviant RVs
were discarded in the orbit  calculation.  In H15, the authors adopted
the SB elements by Duquennoy and fitted only the remaining elements to
the outer orbit.  However, Duquennoy fixed the period of A,B using its
older  visual orbit,  not spectroscopy.   The combined  orbit computed
here uses all the data and removes this inconsistency.

The ``primary''  components are Ba and  Bb, the ``tertiary''  is A (to
get the  orbit of B around  A, change the  outer elements $\Omega_{\rm
  A}$ and  $\omega_{\rm A}$ d by $180^\circ$),  the wobble coefficient
$f =  -0.33$ is negative.  In  the last iteration, we  fixed the outer
period and used only the speckle data for A,B in fitting the remaining
19 free parameters.  As  can be seen in Figures~\ref{fig:111805in} and
\ref{fig:111805out}, the RV curves are  rather noisy owing to the line
blending.   The weighted  rms RV  residuals  are 0.97,  3.5, and  0.78
km~s$^{-1}$  for  Ba, Bb,  and  A,  respectively  (with some  outliers
removed or given  low weight).  Only two uncertain  measures of RV(Bb)
by Duquennoy define  the inner amplitude $K_2$.  The  rms residuals of
the positional  measures are  from 3  to 5 mas  in X  and Y,  for both
orbits.

The {\it  Hipparcos} parallax  of 26.2$\pm$0.6 mas  gives a  too small
mass sum of 2.3 ${\cal M}_\odot$ for the well-defined outer orbit.  We
adopt  the orbital  parallax of  24.1\,mas derived  from  the combined
orbit of  A,B and the mass  sum of 3.0 ${\cal  M}_\odot$, in agreement
with the model masses.   The mass sum in the inner orbit
is 1.82  ${\cal M}_\odot$ and, by  subtraction, the mass of  A is 1.18
${\cal M}_\odot$.  It  matches the inner RV amplitudes,  but the large
error of  $K_2$ makes  the $M  \sin^3 i$ estimate  in the  inner orbit
quite uncertain.

H15 derived  the masses of  A, Ba, and  Bb as 1.12, 0.92,  0.77 ${\cal
  M}_\odot$  using   standard  relations  and   disregarding  the  low
metallicity.  The corresponding spectral  types are F9V, G5V, and K1V.
We repeated  the modeling assuming the orbital  parallax of 24.1\,mas.
The relative photometry  is $\Delta V_{\rm A,B} =  0.50$ mag, based on
the  {\it  Hipparcos} datum  and  some  measurements  by Horch,  while
$\Delta V_{\rm  Ba,Bb} = 1.27$ mag  is measured by Balega  et al.  The
derived masses are 1.14, 1.03,  0.85 ${\cal M}_\odot$, or the mass sum
of 3.02 ${\cal M}_\odot$.  The  combined $K$ magnitude of the model is
5.25,  the observed one  is 5.31  mag. The  model matches  both orbits
quite well.

The model implies $q_{\rm Ba,Bb}  = 0.83$.  The measured wobble factor
$f^* =  -0.33$ (most measures of  A,B refer to the  photo-center of B)
corresponds to  $q_{\rm Ba,Bb} \approx 0.60$, and  the uncertain inner
spectroscopic  mass  ratio is  $q_{\rm  Ba,Bb}  \approx  0.68$.  It  is
possible  that the  component  Bb  is less  massive  and fainter  than
deduced from the photometric model.

\section{Summary and discussion}
\label{sec:sum}

\begin{deluxetable}{l cc cc}
\tabletypesize{\scriptsize}
\tablewidth{0pt}
\tablecaption{Relative orbit orientation and period ratio     \label{tab:phi}}
\tablehead{
\colhead{HIP} &
\colhead{ $P_{\rm out}$} & 
\colhead{ $e_{\rm out}$} & 
\colhead{$P_{\rm out}/P_{\rm in}$} & 
\colhead{$\Phi$}   \\
 & 
\colhead{(yr)} &  &  & 
\colhead{(degrees)} 
}
\startdata
2643   & 70.34 & 0.33 & 14.43$\pm$0.28  & 25.4$\pm$8.5 \\
2643   & 4.85  & 0.14 & 17.57$\pm$0.07  & \ldots \\ 
101955 & 38.68 & 0.12  &  15.41$\pm$0.13  & 64.8$\pm$1.4  \\
103987 & 19.20 & 0.17  &  18.55$\pm$0.08 & 6.2$\pm$9.0 \\
111805 & 30.13 & 0.32  & 20.07$\pm$0.02 & 2.5$\pm$1.5  
\enddata
\end{deluxetable}

We determined  inner and outer  orbits in four multiple  systems using
both  resolved measures and  RVs.  The  ascending nodes  are therefore
identified  without  ambiguity, allowing  us  to  calculate the  angle
$\Phi$ between  the orbital  angular momentum vectors.   These angles,
period ratios,  outer periods  $P_{\rm out}$ and  outer eccentricities
$e_{\rm out}$ are listed in Table~\ref{tab:phi}.  In three systems the
orbits  are  approximately  co-aligned,   and  both  inner  and  outer
eccentricities  are small.   In such  case, the  inner  orbits precess
around  the total  angular momentum,  with $\Phi$  being approximately
constant.    Only   in   HIP~101955   the   orbits   are   closer   to
perpendicularity  than  to  alignment,  $\Phi =  65^\circ$.   In  this
configuration, the  angle $\Phi$ and the  inner eccentricity oscillate
in the  co-called Kozai-Lidov cycles.  Indeed,  the inner eccentricity
in HIP~101955 is  large, $e = 0.61$.  None of  the four close multiple
systems are conter-rotating (all have  $\Phi < 90\degr$), in line with
the general  trend of orbit  co-alignment noted by  \citet{ST02}.  The
massive  counter-rotating close  triple $\sigma$  Ori with  $\Phi \sim
120\degr$ \citep{Schaefer2016} could be formed by a different process.

\begin{figure}
\epsscale{1.0}
\plotone{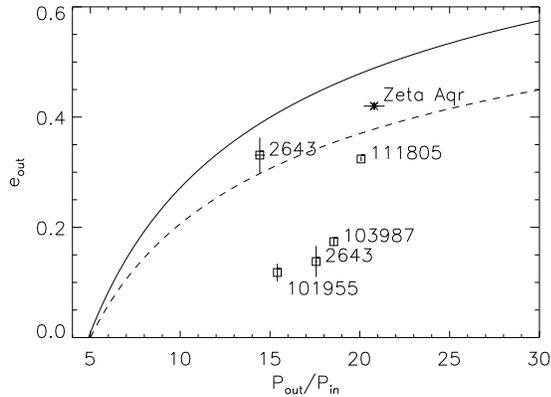}
\caption{Period ratio and outer eccentricity. The full and dashed
  lines are the dynamical stability criteria by \citet{MA2001} and
  \citet{Tok2004}, respectively.
\label{fig:pratio}  }
\end{figure}

Figure~\ref{fig:pratio}   compares  the   period   ratios  and   outer
eccentricities of the multiple systems studied here with the dynamical
stability criterion  of \citet{MA2001}.  The outer  orbit in HIP~2643,
as well as $\zeta$~Aqr \citep{ZetaAqr}, do not satisfy the more strict
empirical criterion of \citet{Tok2004}, which therefore is not valid.

The quadruple  system HIP~2643 with  a 3+1 architecture  resembles the
``planetary'' quadruple HD~91962 \citep{Planetary} in several ways. In
both multiple systems, all  three orbits have moderate eccentricities,
the  outer and middle  orbits are  not far  from coplanarity,  and the
period  ratios  between  the  hierarchical  levels  are  small.   This
suggests that there was some  interaction between the orbits, at least
during the formation  of these systems. However, the  ratio of the two
inner periods in HD~91962 is 19.0, suggesting a mean motion resonance,
while it is not integer in HIP~2643.

\acknowledgements Some  data used  in this work  were obtained  at the
Southern Astrophysical  Research (SOAR) telescope.   We thank E.~Horch
for   critical  re-evaluation   of   the  Gemini   speckle  data   and
communication of  his unpublished observations of HIP  103987.  We
  also thank  both Referees for  careful and inquisitive check  of the
  manuscript.

This work  used the  SIMBAD service operated  by Centre  des Donn\'ees
Stellaires  (Strasbourg, France),  bibliographic  references from  the
Astrophysics Data System maintained  by SAO/NASA, data products of the
Two  Micron All-Sky  Survey (2MASS),  and the  Washington  Double Star
Catalog maintained at USNO.

\facility{Facilities: 
ORO:Wyeth (CfA Digital Speedometer), ​FLWO:1.5m
​ (CfA Digital Speedometer, TRES)​,  SOAR (HRcam), CTIO:1.5m (CHIRON)}


\begin{thebibliography}{99}


\bibitem[Albrecht et al.(2014)]{Albrecht2014}
Albrecht, S., Winn, J. N., Torres, G. et al. 2014, ApJ,  785, 83

\bibitem[Balega et al.(2002)]{Balega2002}
Balega, I. I., Balega, Y. Y., Hofmann, K.-H. et al. 2002, A\&A, 385, 87




\bibitem[Duquennoy(1987)]{D87}
Duquennoy, A. 1987, A\&A, 178, 114 (D87)

\bibitem[Fabrycky et al.(2014)]{Fabrycky2014}
Fabrycky, D. C.,  Lissauer, J. J., Ragozzine, D. et al. 2014, ApJ, 790, 146

\bibitem[Jancart et al.(2005)]{Jancart2005}
Jancart, S., Jorissen, A., Babusiaux, C. \& Pourbaix, D. 2005, A\&A, 442, 365

\bibitem[Hartkopf et al.(2001)]{VB6}
Hartkopf, W. I., Mason, B. D. \& Worley, C. E. 2001, AJ, 122, 3472 (VB6)


\bibitem[Horch et al.(2015)]{H15}
Horch, E. P., van Altena, W. F., Demarque, P. et al. 2015, AJ, 149, 151 (H15)

\bibitem[Kervella et al.(2013)]{Kervella2013}
Kervella, P., M\'erand, A., Petr-Gotzens, M. G. et al. 2013, A\&A, 552, 18

\bibitem[Latham(1992)]{Latham1992}
Latham, D. W. 1992, in ASP Conf. Ser. 32, Complementary Approaches to
Binary and Multiple Star Research, ed. H. McAlister \& W. Hartkopf
(IAU Colloq. 135) (San Francisco: ASP), 110

\bibitem[Latham(1985)]{Latham1985}
Latham, D. W. 1985, in IAU Colloq. 88, Stellar Radial Velocities, ed.
A. G. D. Philip \& D.W. Latham (Schenectady: L. Davis), 21

\bibitem[Latham et al.(2002)]{Latham2002}
Latham, D. W., Stefanik, R. P., Torres, G. et al. 2002, AJ, 124, 1144

\bibitem[Malogolovets et al.(2007)]{Malogolovets2007}
Malogolovets, E. V., Balega, Yu. Yu., \& Rastegaev, D. A. 2007, AstBu, 62, 111

\bibitem[Mason et al.(2001)]{Mason2001}
Mason, B., Hartkopf, W. I., Holdenried, E. R. \& Rafferty, T. J.  2001, AJ, 121, 3224

\bibitem[Mason \& Hartkopf(2014)]{Msn2014a}
Mason, B. D. \& Hartkopf, W. I. 2014, IAUDS, 183, 1

\bibitem[Mardling \& Aarseth(2001)]{MA2001}
Mardling, R. A. \& Aarseth, S. J. 2001, MNRAS, 321, 398

\bibitem[McArthur et al.(2010)]{McArthur2010}
McArthur, B. E., Benedict, G. F., Barnes, R. et al. 2010, ApJ, 715, 1203


\bibitem[Riddle et al.(2015)]{RAO}
Riddle, R. L., Tokovinin, A., Mason, B. D. et al. 2015, ApJ, 799, 4

\bibitem[Schaefer et al.(2016)]{Schaefer2016}
Schaefer, G. H., Hummel, C. A., Gies, D. R. et al. 2016, AJ, 252, 213

\bibitem[S\"oderhjelm(1999)]{Soderhjelm1999}
S\"oderhjelm, S. 1999, A\&A 341, 121 

\bibitem[Sterzik \& Tokovinin(2002)]{ST02}
Sterzik, M. \& Tokovinin, A. 2002, A\&A,  384,  1030

\bibitem[Szentgyorgyi \& Fur\'esz(2007)]{TRES}  
Szentgyorgyi, A.  H., \&  Fur\'esz, G.  2007,  in The  3rd Mexico-Korea
Conference  on Astrophysics: Telescopes  of the  Future and  San Pedro
M\'artir,   ed.  S.    Kurtz,  RMxAC, 28, 129


\bibitem[Tokovinin \& Smekhov(2002)]{TS02}
Tokovinin, A. A. \& Smekhov, M. G. 2002, A\&A, 382,  118

\bibitem[Tokovinin(2004)]{Tok2004}
Tokovinin, A. 2004, RMxAC, 21, 7

\bibitem[Tokovinin et al.(2013)]{CHIRON}
Tokovinin, A., Fischer, D. A., Bonati, M. et al. 2013, PASP, 125, 1336

\bibitem[Tokovinin(2014)]{FG14}
Tokovinin, A. 2014, AJ, 2014, 147, 86

\bibitem[Tokovinin et al.(2015)]{Planetary}
Tokovinin, A., Latham, D. W., \& Mason, B. D.  2015, AJ, 149, 195

\bibitem[Tokovinin et al.(2016)]{SOAR15}
Tokovinin, A., Mason, B.D., Hartkopf, W.I. et al.  2016, AJ, 151, 153

\bibitem[Tokovinin(2016a)]{CHIRON-1}
Tokovinin, A. 2016a, AJ, 152, 11

\bibitem[Tokovinin(2016b)]{ZetaAqr}
Tokovinin, A. 2016b, ApJ, 831, 151 


\bibitem[van Leeuwen(2007)] {HIP2}
van Leeuwen, F. 2007, A\&A, 474, 653

\bibitem[Xu et al.(2015)]{Xu2015}
Xu,   X.-B., Xia, F., \& Fu, Y.-N. 2015, RAA, 15, 1857 

\bibitem[Zhu et al.(2016)]{Zhu2016}
Zhu, L.-Y., Zhou, X., Hu, J.-Y. et al. 2016, AJ, 151, 107 






\end{thebibliography}
\end{document}